# Correct nonlinearity and hysteresis of volt-ampere characteristics of spin valves, magnetic tunnel junctions and memristors

E. S. Demidov

*Nizhniy Novgorod state university, Nizhniy Novgorod, Russia*
demidov@phys.unn.ru

There are essential achievements in synthesis of interesting for creation of compact electronic memory switched by own current structures of spin valves and magnetic tunnel junctions with hysteretic current dependences of resistance. In the offered message the attention to discrepancy to physical principles of a hysteresis of resistance represented in publications is paid. It is schematically presented how the dependences of resistance on current should look not contradicting the energy conservation law for hysteresis dependence of resistance on current and corresponding volt-ampere characteristic.

Magnetic structures switched by own current with hysteresis volt-ampere characteristics (VAC) are interesting by possibility of creation compact magnetoresistive memory in which, unlike systems with operated by external magnetic field, there is no need in additional current write lines with magnetic field slow decaying logarithmically with distance[1,2]. It is easy to be convinced, that for magnetic reversal by a magnetic field of own usual current, even rather for magnetosoft ferromagnetic metal, the density of a current unacceptable on size $j$ from above $10^8$ A/cm$^2$ is required. Essential progress in this plan was outlined in connection with idea of injection of spin polarized carriers of current in structures of the spin valve (SV) – ferromagnetic / not magnetic conductor/ ferromagnetic or magnetic tunnel junction (MTJ) – ferromagnetic / dielectric / ferromagnetic [3,4]. The effect of switching parallel-antiparallel orientations of magnetizations of the magnetic layers divided by not magnetic layer and, hence, magnetoresistance of such structure at change of current direction has been predicted by L. Berger [5] and J.C. Slonczewski [6] and for the first time was observed by J.A. Katine et al [7]. In the first experiments with metal structures Co/Cu/Co for magnetization switching by current too high density of a current through structure ~ $10^7$ A/см$^2$ was required. In [8] it was possible to authors of work to lower on two order (to ~ $10^5$ A/cm$^2$) density of current switching in semiconductor tunnel structure GaMnAs/InGaAs/GaMnAs at helium temperatures. Recently there were messages about structures of SV [9] and MTJ [10] with Heusler alloys magnetic layers for which already at a room temperature the hysteretic dependences of resistance on current $R(I)$ with lowered to $10^6$ A/cm$^2$ current density of the switching were observed. In present article the attention to a physical incorrectness resulted in articles [7-10] of presentations of current hysteretic dependences of differential resistance $R(I)$ for self-switched magnetic structures is paid. It is presented how the hysteresis dependence of resistance of structure on a current and corresponding volt-ampere characteristics should look not contradicting the law of conservation of energy.

Various variants of mathematical description of a hysteresis with simple parametrical representation by harmonious functions or complicated integrating descriptions are known[11]. The variant simple, evident and more convenient for computer modeling is offered here. Resulted in the above mentioned publications dependence $R(I) = f(I)$ look like, schematically presented in Fig. 1a, here is simulated by defined in arbitrary units on axis $I$ with step $\Delta I$ on the way ($I_{min} \rightarrow I_{max} \rightarrow I_{min}$) discrete function

$$f(I)=f_0/(1+\exp(\alpha(-\mathrm{Sign}(\Delta I)\beta-I)))+\varphi, \quad (1)$$

where $f_0=1$, $\varphi = 1$ - together with $f_0$ defines hysteresis degree, $\alpha = 10$ - sharpness of hysteresis, $\beta = 10$ - « coercitive force» defines width of loop of hysteresis, Sign (x) =1 at x> 0 and Sign (x) =-1 at x <0. Corresponding to this dependence $R(I)$ the VAC, turning out by integration of $R(I)$ in Fig. 1a with repetition of cycles ($I_{min} \rightarrow I_{max} \rightarrow I_{min} \rightarrow \ldots$) of consecutive passage of hysteresis loop, is, apparently in Fig. 1b, divergent and contradicting the energy conservation law. Correct hysteresis resistive dependences $R(I)$ of self-switched structure should contain loops compensating each other at integration with passage in opposite directions as it is schematically shown in Fig 1c. Such dependence $R(I)$ is modeled by a linear combination of function (1) and its first discrete derivative $\Delta f/\Delta I$

$$R(I)=f-(\Delta f/\Delta I)\gamma, \quad (2)$$

where providing compensating of loops of hysteresis the multiplier is presented in form

$$\gamma=(\beta+|\Delta I/2|)\mathrm{Sign}(\Delta I). \quad (3)$$

Corresponding Fig. 1c VAC with a hysteresis looks like, shown in Fig. 1d and is defined numerical by cyclic integration ($I_{min} \rightarrow I_{max} \rightarrow I_{min}$) in borders $I_{max} = - I_{min}=20$

$$V(I)=\sum_{I_{min}}^{I} R(i)\Delta i+C, \quad (4)$$

where $I_{min} <0$, $C = (I_{min} - |\Delta I|) \varphi$ - provides $V=0$ at $I=0$. It is important to consider, that resistive structures with self-switching except nonlinearity and a hysteresis necessarily should be asymmetric as it is visible in Fig. 1d.

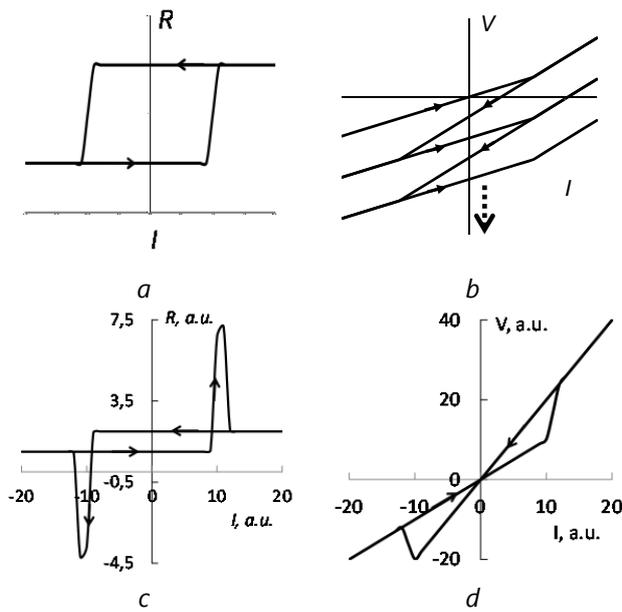

Fig. 1. Schematically view of *a* - current dependence of differential resistance *R(I)* for self-switching magnetic structures in [7-10], *b* - corresponding this dependence divergent VAC, *c* - correct dependence *R(I)* for self-switched structure with parameters $f_0=1$, $\varphi=1$, $\alpha=10$, $\beta=10$ and *d* - corresponding Fig. 1*c* view of VAC with hysteresis.

Possibly at measurements by authors of [7-10] of electrotransport characteristics of layered structures the inertial equipment smoothing sharp spikes of *R(I)* around switching of mutual orientation of magnetisation of magnetic layers was applied. Therefore, to avoid hardware distortions at use of precision inertial measuring technics, probably, it is expedient to record directly VAC, which then numerically to differentiate for definition of difference of differential resistance at direct and return transiting of VAC.

It is remarkable, that dependence *R(I)* in Fig. 1*c* and corresponding to it VAC in Fig.1*d* contains the site with negative differential resistance and, hence, possibility of construction of the amplifier or the generator of electromagnetic oscillations. The realizability of such phenomenon requires separate experimental and theoretical research. Here we notice that apparently at a certain set of parameters the variant without the site with negative *R(I)*, as is shown in Fig. 2 is possible. The following set of parametres is used: $f_0=1$, $\varphi=3$, $\alpha=1$, $\beta=10$. At weak display of a hysteresis there can be useful plotting of deviation $\Delta V = V(I) - V_L(I)$ (Fig.2*c*) from linear dependence $V_L(I) = IR_0$, shown in Fig. 2*b* by dotted line.

In the conclusion we will notice, that the stated qualitative reasons should be applicable and for the self-switched structures based on other not magnetic physical phenomena, for example, atomic current transport in memristors [12]. Certainly, in Fig. 1, 2 the refined, reflecting main features of display of a hysteresis, are resulted. It is obvious, that real structures will differ by additional curvatures *R(I)* or *V(I)*, caused, for example, by temperature dependence of resistance because of current heating.

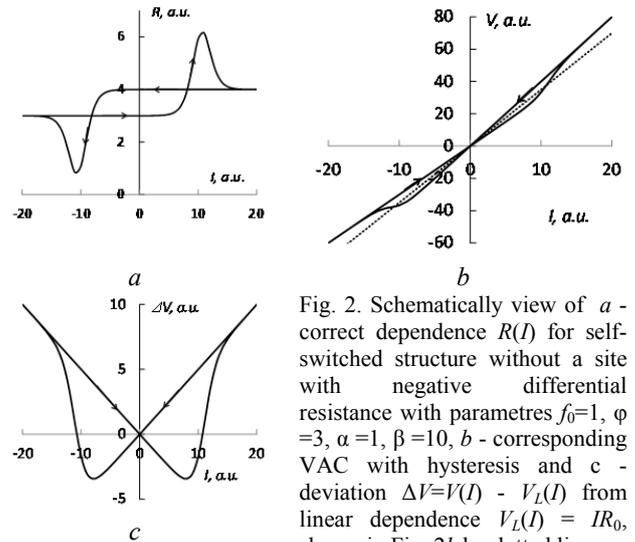

Fig. 2. Schematically view of *a* - correct dependence *R(I)* for self-switched structure without a site with negative differential resistance with parametres $f_0=1$, $\varphi=3$, $\alpha=1$, $\beta=10$, *b* - corresponding VAC with hysteresis and *c* - deviation $\Delta V = V(I) - V_L(I)$ from linear dependence $V_L(I) = IR_0$, shown in Fig. 2*b* by dotted line.

This study was supported by grants of Russian Foundation for Basic Research (project nos. 05.02.17362, 08.02.01222a, and 11.02.00855a), ISTC G1335, Analytical department purpose program of Higher school 2.1.1/2833, and 2.1.1/12029, and the Ministry of Education and Science of the Russian Federation (contract no. 02.740.11.0672).